\documentclass[a4paper]{jpconf}
\usepackage{graphicx}
\usepackage{subfigure}
\begin{document}
\title{First principles perspective on the microscopic model for Cs$_2$CuCl$_4$ and Cs$_2$CuBr$_4$}

\author{K Foyevtsova, Y Zhang,  H O Jeschke and R Valent\'{\i}}

\address{Institut f\"ur Theoretische Physik,
Goethe-Universit\"at Frankfurt, Max-von-Laue-Stra{\ss}e 1, 60438
Frankfurt am Main, Germany}

\ead{foyevtsova@itp.uni-frankfurt.de}

\begin{abstract}
   We investigate the microscopic model for the frustrated layered
  antiferromagnets Cs$_2$CuCl$_4$ and Cs$_2$CuBr$_4$ by performing
  {\it ab initio} density functional theory (DFT) calculations and
  with the help of the tight-binding method.  The combination of both
  methods provide the relevant interaction paths in these materials,
  and we estimate the corresponding exchange constants.  We find for
  Cs$_2$CuCl$_4$ that the calculated ratio of the strongest in-plane
  exchange constants $J'/J$ between the spin-1/2 Cu ions agrees well
  with neutron scattering experiments. 
  The microscopic model based on the derived exchange constants is
  tested by comparing the magnetic susceptibilities obtained from
  exact diagonalization with experimental data.  The electronic
  structure differences between Cs$_2$CuCl$_4$ and Cs$_2$CuBr$_4$ are
  also analyzed.
\end{abstract}

\section{Introduction}
The frustrated layered antiferromagnets Cs$_2$CuCl$_4$ and
Cs$_2$CuBr$_4$ have recently attracted a lot of interest because of
unconventional magnetic properties \cite{J_Br, frac, BEC}. While in
Cs$_2$CuCl$_4$, a magnetic field induced Bose-Einstein condensation
(BEC) of magnons \cite{BEC} and spin-fractionalization \cite{frac}
have been observed, Cs$_2$CuBr$_4$, though being iso-structural to
Cs$_2$CuCl$_4$, shows no sign of BEC. Instead, its field-dependent
magnetization shows two plateaux \cite{J_Br}.

Both compounds have been classified as two-dimensional systems, with
an underlying anisotropic triangular lattice of spin-$1/2$ Cu$^{2+}$
ions (right panel of figure \ref{uCell}).  The ratio of the two
antiferromagnetic exchange constants in the triangular lattice, $J'/J$
(see figure \ref{uCell}) has been found from neutron scattering
experiments to be 0.34 for Cs$_2$CuCl$_4$ \cite{J_Cl} and 0.74 for
Cs$_2$CuBr$_4$ \cite{J_Br}.  The interlayer couplings as well as the
Dzyaloshinski-Moriya interaction also present in these systems are
suggested to be an order of magnitude smaller than $J$ and $J'$.

Yet, since extracting the exchange constants from the neutron
scattering experiment data is based on the assumption of a suitable
Hamiltonian, there is a need to support experimental findings with
theoretical predictions.  This motivated us to perform density
functional theory calculations on Cs$_2$CuCl$_4$ and Cs$_2$CuBr$_4$ as
presented in this work.

\section{Computational details}
We have performed density-functional calculations in the generalized
gradient approximation (GGA) \cite{GGA}, using the full-potential
linearized augmented plane wave (LAPW) method implemented in the
WIEN2k code \cite{WIEN2k}. With $RK_M=7.0$ and 120 $k$-points in the
irreducible Brillouin zone, we obtained well-converged results for
both Cs$_2$CuCl$_4$ and Cs$_2$CuBr$_4$. For Cs$_2$CuCl$_4$ we took the
experimental room-temperature structure reported in \cite{Cl-str}, and
for Cs$_2$CuBr$_4$ we used the structure reported in \cite{Br-str}.

\section{Crystal structure}
\begin{figure}[h]
\subfigure {\includegraphics[trim = 8mm 17mm 11mm 25mm, clip,width=4 cm]{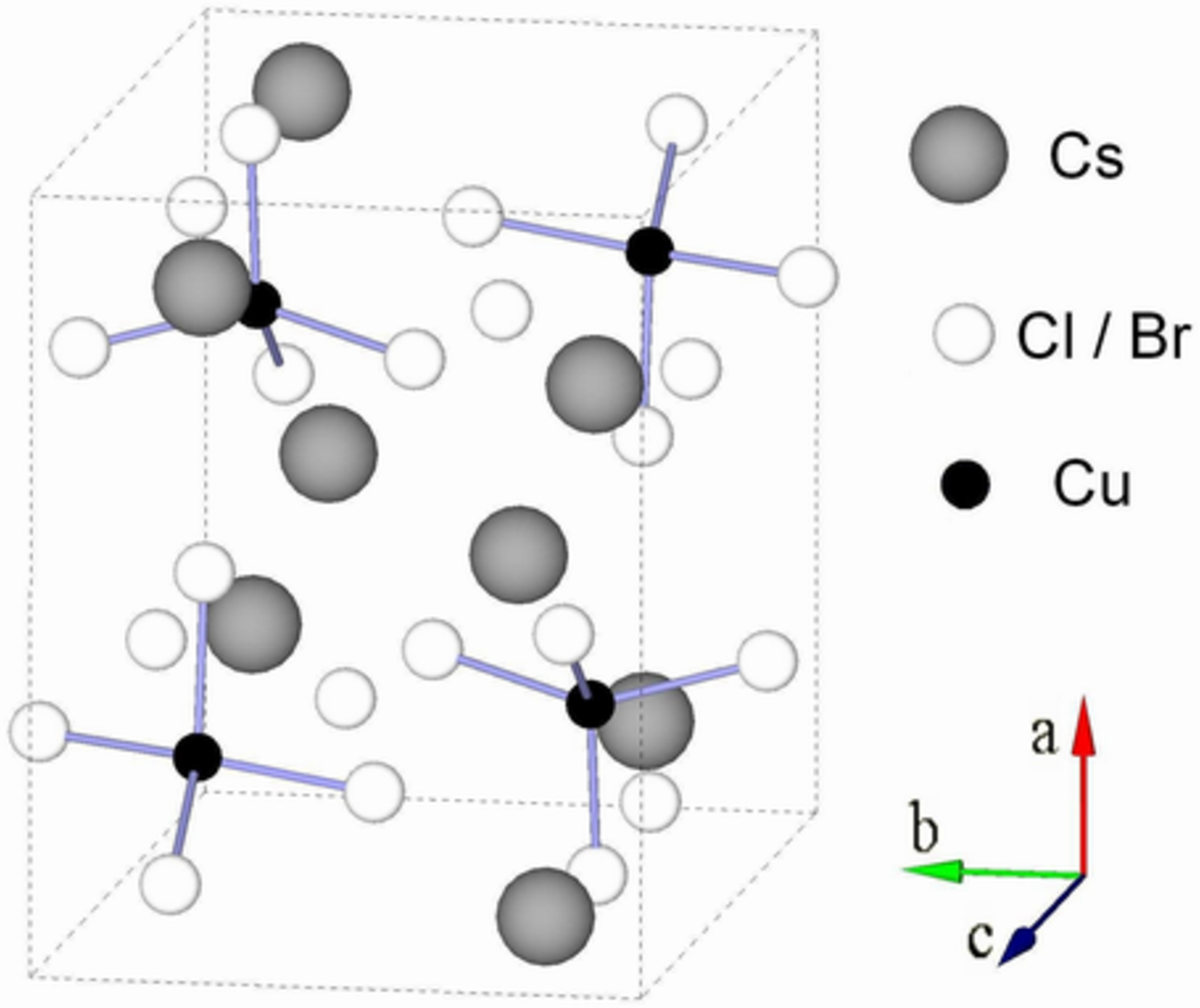}}\hspace{0.8cm}
\subfigure {\includegraphics[trim = 0mm 11mm 0mm 5mm, clip,width=5.0 cm]{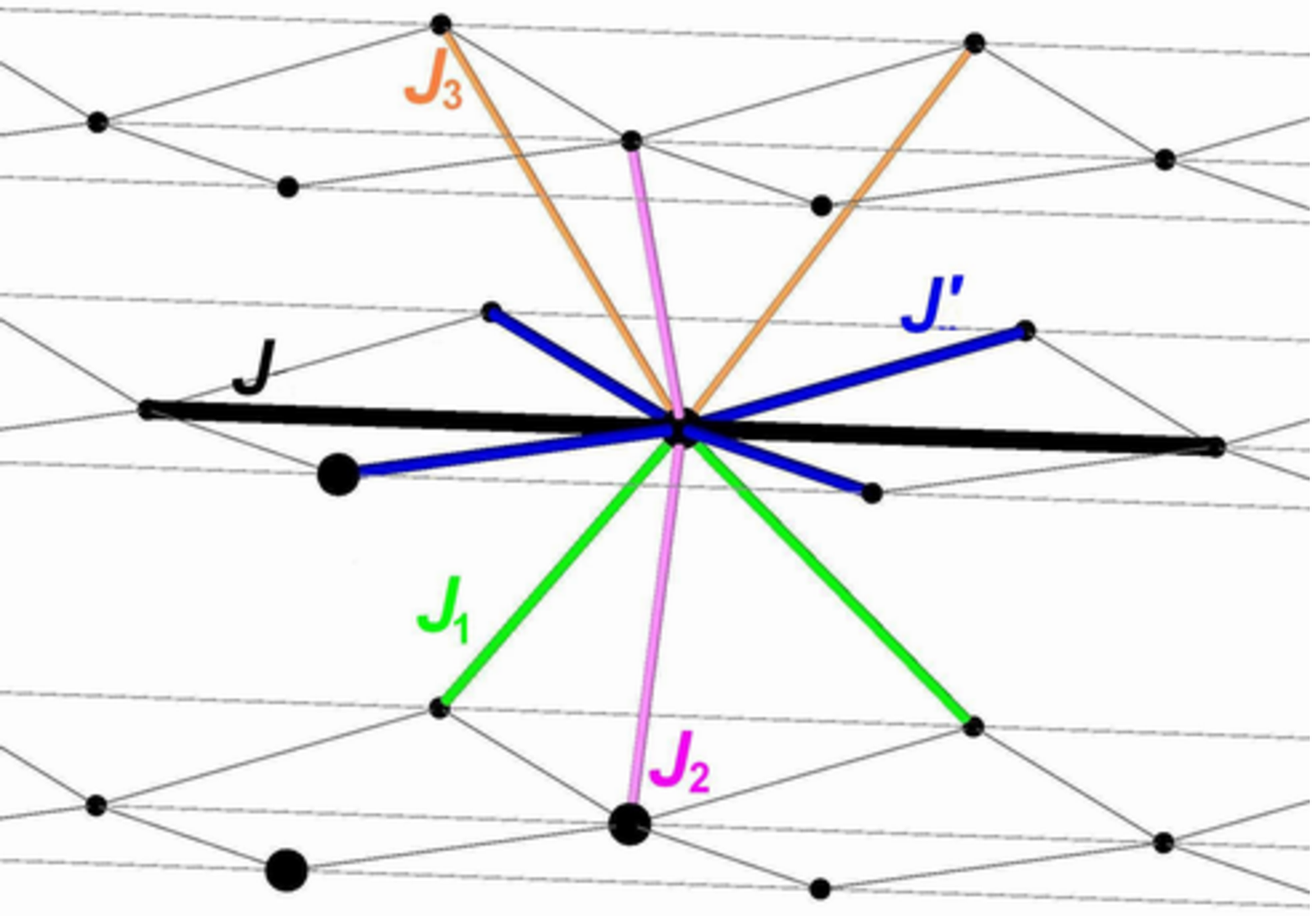}}\hspace{0.8cm}
\begin{minipage}[b]{4.5cm}
\caption{Left: Unit cell of Cs$_2$CuBr$_4$ and Cs$_2$CuCl$_4$. Right:
 Lattice of magnetic Cu sites where the  five most important 
interaction pathways are shown.}
\label{uCell}
\end{minipage}
\end{figure}

Cs$_2$CuCl$_4$ and Cs$_2$CuBr$_4$ crystallize in the orthorhombic
space group \textit{Pnma} (see figure \ref{uCell}). The unit cell
contains four formula units and has lattice parameters $a=9.769$ \AA,
$b=7.607$ \AA\; and $c=12.381$ \AA\; for Cs$_2$CuCl$_4$ and $a=10.195$
\AA, $b=7.965$ \AA\; and $c=12.936$ \AA\; for Cs$_2$CuBr$_4$.  Each of
the four equivalent Cu$^{2+}$ ions is surrounded by four halogen
(Cl$^-$ or Br$^-$) ions located at the vertices of a squeezed
tetrahedron. The CuCl$_4^{2-}$/CuBr$_4^{2-}$ complexes arranged in a
two-dimensional triangular lattice form layers parallel to the $bc$
plane separated along $a$ by Cs$^+$ ions. The most important
superexchange pathways between Cu ions have been determined
\cite{J_Cl, J_Br} to be those within a CuCl$_4^{2-}$/CuBr$_4^{2-}$
layer ($J$ and $J'$ in the right panel of figure \ref{uCell}).  The
magnetic interaction is mediated via halogen ions.

\section{Electronic structure}
\begin{figure}[h]
\subfigure {\includegraphics[width=4.9 cm]{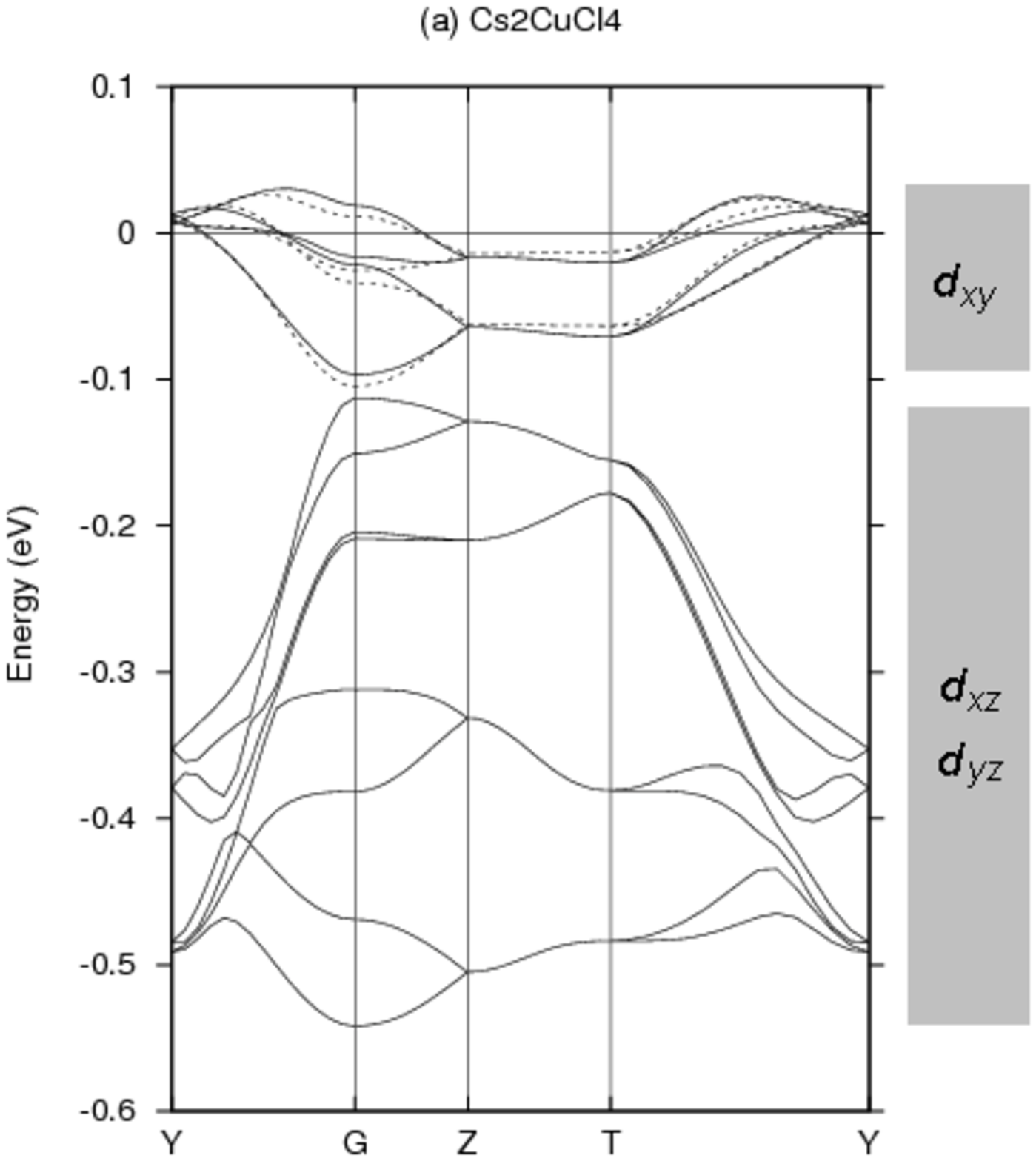}}\hspace{1pc}
\subfigure {\includegraphics[width=4.9 cm]{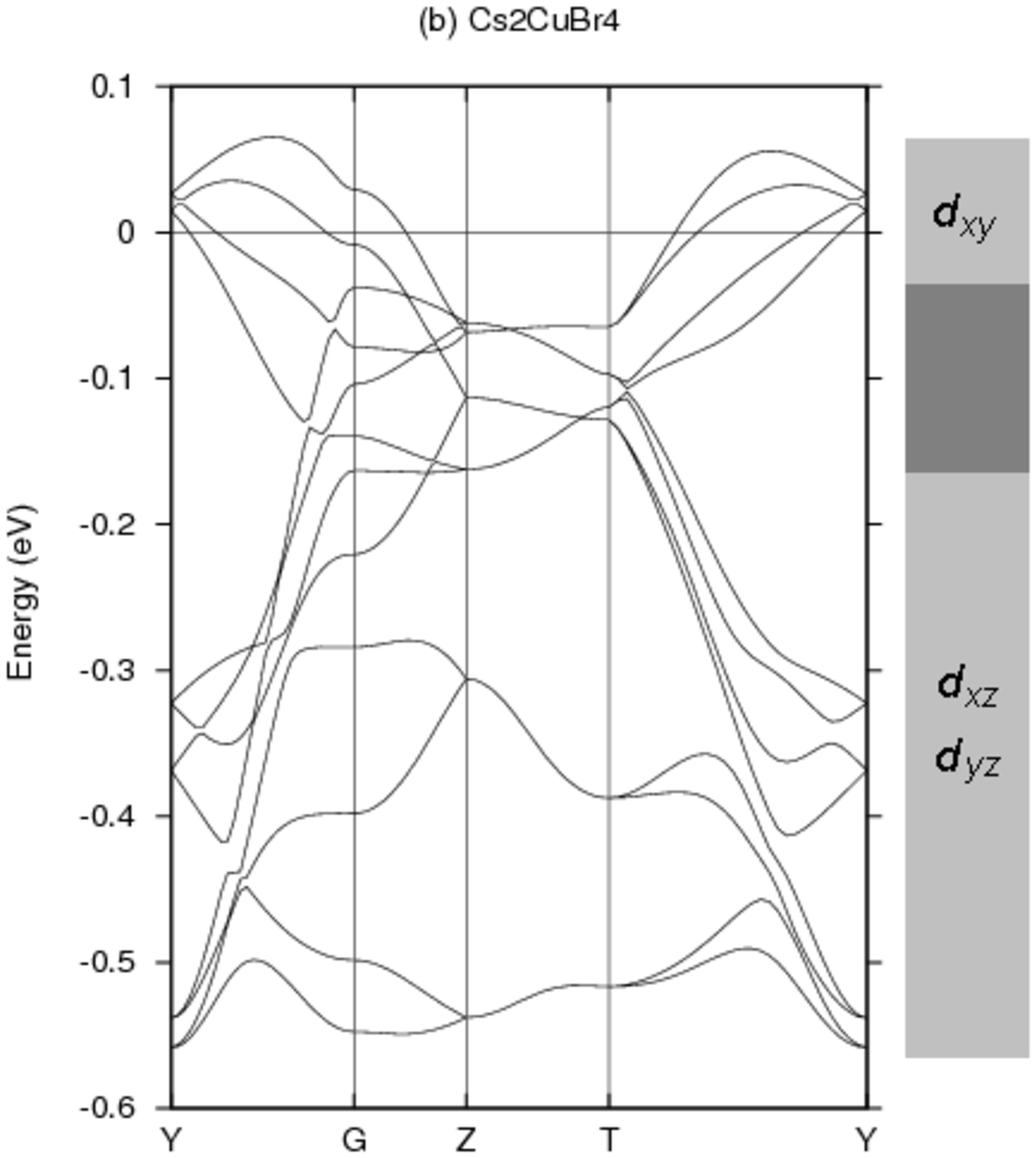}}\hspace{1pc}
\begin{minipage}[b]{5.2cm}\caption{\label{el_str}Bandstructure of
    Cs$_2$CuCl$_4$ (a) and Cs$_2$CuBr$_4$ (b). Dashed lines in (a) are
    the tight-binding fit.}
\end{minipage}
\end{figure}

In panels (a) and (b) of figure \ref{el_str} we present the calculated
bandstructures of Cs$_2$CuCl$_4$ and Cs$_2$CuBr$_4$.
Cu~$t_{2g}$ states that are hybridized via halogen $p$ states lie in the range of
energies between -0.6 eV and 0.1 eV. The twelve bands originate from
the three Cu~$t_{2g}$ orbitals which are split due the distorted
tetrahedron crystal field of halogen ions.  In Cs$_2$CuCl$_4$, the
four partially unoccupied higher-energy Cu~$3d_{xy}$ bands are separated
from the nearly degenerate Cu~$3d_{xz}$ and Cu~$3d_{yz}$ bands by a narrow gap.  In
Cs$_2$CuBr$_4$, the gap is absent. Out of these bandstructure features
we conclude that the low-energy behavior of Cs$_2$CuCl$_4$ can be
described by a one-band model ($d_{xy}$) while the three Cu~$t_{2g}$ band
model is necessary for the description of Cs$_2$CuBr$_4$.

Both Cs$_2$CuCl$_4$ and Cs$_2$CuBr$_4$ are Mott insulators.  The fact
that the GGA functional gives a metallic state is due to the
insufficient treatment in this approach of electronic correlations.
Consideration of GGA+U provides the correct insulating behavior. Since
we are interested in obtaining the effective hopping matrix elements
between the Cu ions, we will consider in what follows the GGA
exchange-correlation functional.

The Cu~$3d_{xy}$ bands of Cs$_2$CuCl$_4$ are comparatively less
dispersive than those of Cs$_2$CuBr$_4$. It is then to be expected
that the hopping parameters between magnetic Cu ions in Cs$_2$CuCl$_4$
will be smaller than in Cs$_2$CuBr$_4$.  In Cs$_2$CuCl$_4$, the bands
are almost flat in the $k_x$-direction, which indicates that layers of
Cu ions (parallel to the $bc$-plane) are weakly coupled.  Both
observations are in agreement with experiment.

\section{Calculation of exchange couplings for Cs$_2$CuCl$_4$}
Based on the LAPW bandstructure (figure \ref{el_str} (a)), we have
calculated hopping integrals between magnetic Cu ions of
Cs$_2$CuCl$_4$ using the tight-binding (TB) method.  We constructed a
single-orbital TB Hamiltonian intended to describe the bands near the
Fermi level, which are the Cu~$3d_{xy}$ bands in Cs$_2$CuCl$_4$. The
hopping integrals $t$ entering the Hamiltonian were varied until the
best possible fit of the TB bands to the LAPW bands was achieved
(dashed lines in figure \ref{el_str} (a)).  Application of the TB
method for Cs$_2$CuBr$_4$ is very involved since one needs to consider
a three-orbital Hamiltonian to describe the overlap of the three
Cu~$t_{2g}$ bands.  The number of variational parameters of the
three-band model Hamiltonian greatly increases due to all possible
inter-orbital hoppings.

In table \ref{t} we list the eight hopping integrals $t$ that we
included in the Hamiltonian for Cs$_2$CuCl$_4$.  The interaction
pathways corresponding to $t$, $t'$, $t_1$, $t_2$ and $t_3$ are shown
in figure \ref{uCell} (labeled by the corresponding exchange constants
$J$).
\begin{table}[h]
  \caption{\label{t}Calculated hopping integrals $t_i$ with corresponding 
    exchange constant ratios $(J_i/J_{\rm max})=(t_i/t_{\rm max})^2$ versus 
    experimentally determined exchange constants $J^{\rm exp}$ for Cs$_2$CuCl$_4$.} 
\begin{center}
\lineup
\begin{tabular}{*{9}{l}}
\br                              
Interaction paths ($p$)  & $p_J$ & $p_{J'}$ & 
$p_{J_1}$ & $ p_{J_2}$ & $p_{J_3}$ & $p_{J_4}$ & 
$p_{J_5}$ & $p_{J_6}$ \cr 
&&&&&&&&\cr
\mr
$t_i$ (meV)&-11.7 &-6.75&4.20&-5.18&-8.39&-2.17&-4.11&-3.35\cr
$(J_i/J_{\rm max})=(t_i/t_{\rm max})^2$& \0 1 &\00.336&0.131&\00.197&\00.516&\00.033 &\00.123&\00.074\cr
$(J_i/J_{\rm max})^{\rm exp}$& \0 1 &\00.342&\m---&\00.045&\m\0---&\m\0--- &\m\0---&\m\0---\cr
\br
\end{tabular}
\end{center}
\end{table}
$J = J^{\rm FM} + J^{\rm AFM}$ where $J^{\rm FM}$ is the ferromagnetic component
of the exchange while $J^{\rm AFM}$ describes the antiferromagnetic
contributions to the total exchange constant. $J^{\rm AFM}_i$ can be
estimated by applying perturbation theory on the Hubbard model as
$J^{\rm AFM}_i=4t^2_i/U$ where $t_i$ are the hopping integrals and $U$ is
the on-site Coulomb repulsion. An alternative way of calculating $J$
is by considering total energies of various magnetic configurations.
In Ref. \cite{unpublished} we found by considering total energy
differences that the ferromagnetic contributions are negligible for
the strongest interactions, which allows us to compare the obtained
$J^{\rm AFM}_i$ from $t_i$ with the $J_i$ obtained from experimental data.
In table \ref{t} we present the calculated (second row) and
experimentally fitted (third row) ratios $J_i/J_{\rm max}$, which are
independent of $U$.  We observe that the calculated in-plane exchange
ratio $J'/J=0.336$ agrees well with the experimentally fitted 0.342.
At the same time, we observe that the antiferromagnetic contribution
of the interlayer exchange constants $J_1$, $J_2$ and $J_3$ is also
considerable, which implies a non-negligible three dimensional
contribution of the exchange interactions in Cs$_2$CuCl$_4$.

\begin{figure}[h]
\subfigure {\includegraphics[trim = 22mm 15mm 65mm 40mm, clip,width=6.5 cm]{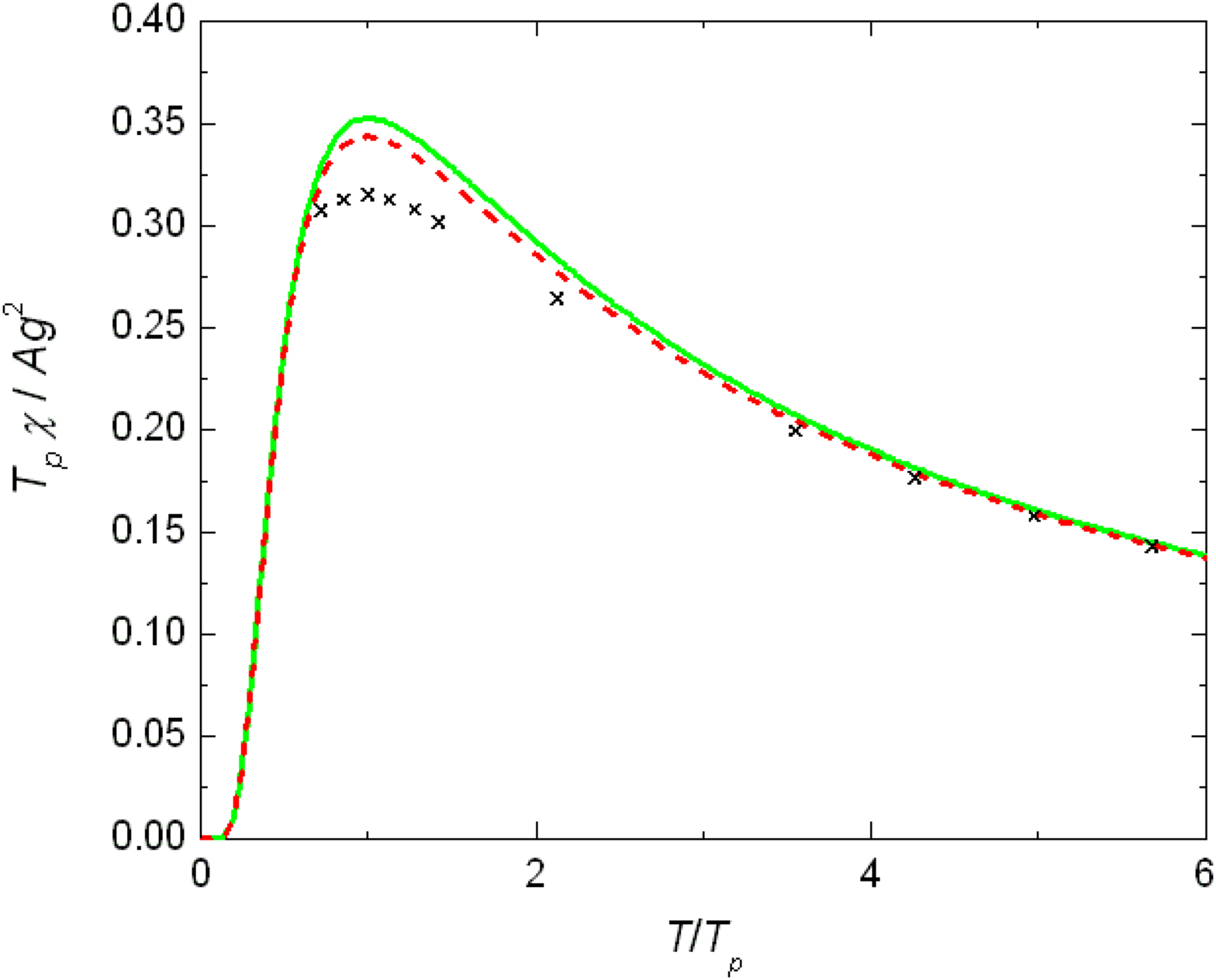}}\hspace{0.5cm}
\begin{minipage}[b]{8.5cm}\caption{\label{susc}Susceptibility of
    Cs$_2$CuCl$_4$ vs. temperature in units of the peak temperature
    $T_p$. Green solid line: DFT-based model with $J$ and $J'$; red
    dashed line: DFT-based model with $J$, $J'$ and $J_3$; crosses:
    experimental data \cite{susc}. The susceptibility is scaled as
    $T_p\chi/Ag^2$ with $A=N_A\mu_B^2/4k=0.0938$ cgs units and a $g$
    factor of $g=2.17$.}
\end{minipage}
\end{figure}

The magnetic susceptibility of a Heisenberg Hamiltonian for
Cs$_2$CuCl$_4$ based on exchange couplings from our DFT calculations
was calculated with the exact diagonalization method (the Hamiltonian
of a 16-site system was diagonalized with periodic boundary
conditions). Two models were investigated: one including $J$ and $J'$
only and the other including $J$, $J'$ and the largest interlayer
coupling $J_3$. In figure \ref{susc} we present the results together
with experimental data.  The susceptibility plot is rescaled in a way
that only relative values of $J$'s play a role \cite{Zheng}.

We observe that inclusion of the interlayer coupling $J_3$ has very
little effect on the susceptibility and therefore other measurements
like neutron scattering along the direction perpendicular to the $bc$
plane are needed to investigate such possible interlayer couplings.
We note that the less than perfect agreement of our exact diagonalization
calculations with experiment is due to finite size
effects\cite{Zheng}.

In summary, we presented electronic structure calculations of
Cs$_2$CuCl$_4$ and Cs$_2$CuBr$_4$.  While Cs$_2$CuCl$_4$ can be well
described by a single-band model with important $J$, $J'$ intralayer
interactions and non-negligible interlayer coupling, a three-band
model is needed for Cs$_2$CuBr$_4$ due to the strong involvement of
all Cu~$t_{2g}$ bands near the Fermi surface. We found that the
magnetic susceptibility for Cs$_2$CuCl$_4$ is not drastically changed
by the inclusion of interlayer couplings in the model and therefore
alternative measurements have to be considered to investigate the
interlayer couplings in this material.

\ack{We gratefully acknowledge support from the Deutsche
  Forschungsgemeinschaft via SFB/TRR 49 and Emmy Noether programs.}

\end{document}